\documentstyle[twoside,fleqn,espcrc2]{article}

\begin{document}

\author{Itzhak Bars \\
Department of Physics and Astronomy\\
University of Southern California\\
Los Angeles, CA 90089-0484}
\title{{\small \noindent hep-th/9601164 \hfill USC-96/HEP-B1}\bigskip\\
Consistency between 11D and U-duality\thanks{%
Based on lectures delivered at the {\it 29th Int. Symp. Ahrenshoop on the
Theory of Elementary Particles, }July 1995, Buckow, Germany; and at {\it %
Strings, Gravity and Physics at the Planck Scale, }Aug. 1995{\it , }Erice,
Italy.}}

\begin{abstract}
U duality transformations must act on a basis of states that form complete
multiplets of the U group, at any coupling, even though the states may not
be mass degenerate, as for a broken symmetry. Similarly, if superstring
theory is related to a non-perturbative 11D M-theory, then an 11D
supermultiplet structure is expected, even though the multiplet may contain
states of different masses. We analyse the consistency between these two
multiplet schemes at the higher excited string levels for various
compactifications of the type IIA superstring. While we find complete
consistency for a number of compactifications, there remain some unanswered
questions in others. The relation to D-branes also needs further
clarification.
\end{abstract}

\maketitle

\section{Introduction}

Circumstancial evidence for U-duality in string theory has been accumulating 
\cite{ht}-\cite{circumstancial2}. Also, starting with 11D supergravity in
the form of the low energy limit, there seems to be hints \cite{witten} for
an 11 dimensional extended structure lurking behind the non-perturbative
aspects of string theory \cite{ib11d}\cite{townsend}\cite{jhs2}. Further
study of duality and the possible connections with 11-dimensions is bound to
reveal more non-perturbative properties of the underlying theory. In this
lecture I will concentrate on the assumed 11D theory and will try to discuss
parts of its stringy spectrum and its consistency with U-duality. In
particular, I will examine the toroidally compactified type-II superstring
on $R^d\times T^c$ (with $d+c=10)$ and analyse its spectrum for various
values of $\left( d,c\right) $, looking for connections to 11D and
U-duality. This is a summary of the work in \cite{ib11d}\cite{ibyank} and
more recent observations along the same lines regarding BPS saturated
stringy states which are related to D-branes.

Beginning with the uncompactified type IIA string theory (d=10, c=0), in 
\cite{ib11d} it was suggested that, if the type IIA string theory on $R^{10}$
is related to a non-perturbative 11D structure, then its stringy spectrum
(at any strength of the coupling) must display a certain pattern consistent
with 11D supersymmetry and 11D Lorentz group. The patterns were displayed up
to excitation level $l=5$ as follows.

\subsection{Hidden 11D in 10D superstring}

In the Green-Schwarz formalism, the perturbative states of type IIA
superstring in 10D, at level $l$%
\begin{equation}
({\rm {Bose}\oplus {Fermi oscillators})^{\left( l\right) }|vac,\,\,p^\mu >}
\end{equation}
start out as SO(8) supermultiplets in the light-cone gauge. Here $vac$
stands for the $2_B^7+2_F^7$ dimensional Clifford vaccum of 8-left and
8-right fermionic zero modes. Because of Lorentz invariance in the critical
dimension, one expects that massless states must form SO(8) supermultiplets,
while massive states must form SO(9) supermultiplets. However, by detailed
examination of the oscillator states one finds that there are higher
structures that begin to give more hints of 11-dimensions. The oldest of
these hints is that the massless sector comes as SO(9) multiplets ($%
l=0:2_B^7+2_F^7,\,$ identical to the massless states of 11D supergravity),
while the first massive level comes as SO(10) supermultiplets \cite{ibmembr}%
\cite{ib11d} 
\begin{equation}
l=1:\quad 2_B^{15}+2_F^{15}.
\end{equation}
In fact, these correspond to the short and long multiplets of 11D
supersymmetry and therefore are expected to occur in any supersymmetric
theory that has 11D supersymmetry. The fact that they arise in the 10D
string theory is already of striking significance. This should be viewed
partly as a consequence of the fact that the 32 components of the two 10D
supercharges of type IIA reassemble to a single supercharge in 11D, and
partly due to the oscillator content of the theory.

The states above are purely perturbative. In \cite{witten} it was suggested
that the non-perturbative BPS saturated ``black hole'' solutions of \cite{ht}
can be interpreted as Kaluza-Klein excitations of 11D supergravity, thus
elevating the $l=0,$ $2_B^7+2_F^7,$ multiplet from being 10-dimensional
fields $\phi \left( x^\mu \right) $ to 11-dimensional fields $\phi \left(
x^\mu ,x^{11}\right) .$ From the point of view of the vacuum state this
corresponds to having a quantized 11th momentum 
\begin{equation}
|vac,\,\,p^\mu ,p^{11}>
\end{equation}
which is the central extension in the type IIA 10D supersymmetry algebra,
that occurs non-perturbatively in the 11D superalgebra. It is then natural
to expect that, in the theory on $R^{10}$ (or in its assumed
non-perturbative 11D structure), there are non-perturbative string
excitations with non-trivial values of the central extension {\it at all
excitation levels, at any coupling} 
\begin{equation}
({\rm {Bose}\oplus {Fermi\ oscill.})^{\left( l\right) }|vac,\,\,p^\mu
,p^{11}>.}  \label{osc}
\end{equation}
{\it \ } Thus, the $2_B^{15}+2_F^{15}$ multiplet at the first massive level $%
l=1$ is also elevated to 11-dimensional fields $\phi \left( x^\mu
,x^{11}\right) $. Then the $l=0,1$ fields know about 11 dimensions in two
ways: $\left( i\right) $ the indices, $\left( ii\right) $ the momentum (or
space) dependence. We will refer to these two spaces as ``index space'' and
``base space'' respectively.

Since the $l=0$ states are BPS saturated one can give their mass in the form 
\cite{ht}\cite{witten} 
\begin{equation}
l=0:\quad M_{10}=\left| p_{11}\right| .
\end{equation}
This is equivalent to saying that the 11-dimensional mass is zero for all
the non-perturbative or perturbative states 
\begin{equation}
l=0:\quad M_{11}=\sqrt{M_{10}^2-p_{11}^2}=0.  \label{massp11}
\end{equation}
In \cite{witten} it was pointed out that $M_{10}=\left| p_{11}\right|
=cn/\lambda $ become massless at infinite coupling $\lambda \rightarrow
\infty .\,\,$In \cite{ib11d} it was emphasized that $M_{11}=0$ at all values
of the coupling. Stated in this way all the sates satisfying the condition $%
M_{11}=0$ are viewed as a single multiplet of degenerate states in the
11-dimensional theory, {\it at any coupling. }From the point of view of 10D,
the eleven dimensional symmetry is broken, but {\it its 11D multiplet
structure is preserved}. For the $l=1$ states we cannot give a
non-perturbative mass formula, although we may still {\it define} $M_{11}=%
\sqrt{M_{10}^2-p_{11}^2}\neq 0.$ It is evident that we still have an 11D
multiplet structure for all $l=1$ states, including all those that have
non-perturbative values of $p_{11}$.

The {\it perturbative} index space at higher levels is obtained by examining
the content of the oscillators in (\ref{osc}). For the theory in 10D this is
done explicitly in \cite{ib11d} with the following result 
\begin{equation}
indices\Rightarrow \left( 2_B^{15}+2_F^{15}\right) \times R^{\left( l\right)
}.  \label{10dpert}
\end{equation}
The factor $2_B^{15}+2_F^{15},$ although obtained from a combination of the
oscillators and $vac,$ can be reinterpreted as the action of 32 supercharges
on a set of $SO(9)$ representations $R^{\left( l\right) }$ at oscillator
level $l,$ where $SO(9)$ is the spin group in 10-dimensions for massive
states. The factor $R^{\left( l\right) }$ is of the form of direct products
of $SO(9)$ representations coming from left/right movers 
\begin{equation}
R^{\left( l\right) }=\left( \sum_ir_i^{(l)}\right) _L\times \left(
\sum_ir_i^{(l)}\right) _R
\end{equation}
such that the left-factor is identical to the right-factor\footnote{%
The left/right excitation levels are the same for the 10D theory, $%
l=l_L=l_R. $}, and is given by the collection of $SO(9)_{L,R}$
representations listed in Table 1, where the subscripts $B/F$ stand for
boson/fermion respectively.

\begin{table}[tbp]
\caption{L/R oscillator states in 10D superstring.}
\label{tab: tI}\centering
$
\begin{tabular}{|l|l|}
\hline
Level & SO(9)$_{L,R}$ reps $\left( \sum_ir_i^{(l_{L,R})}\right) _{L,R}$ \\ 
\hline
$l_{L,R}=1\quad $ & $1_B$ \\ \hline
$l_{L,R}=2$ & $9_B$ \\ \hline
$l_{L,R}=3$ & $44_B+16_F$ \\ \hline
$l_{L,R}=4$ & $(9+36+156)_B+128_F$ \\ \hline
$l_{L,R}=5$ & $
\begin{array}{l}
\left( 
\begin{array}{c}
1+36+44+84 \\ 
+231+450
\end{array}
\right) _B \\ 
+\left[ 16+128+576\right] _F
\end{array}
$ \\ \hline
\end{tabular}
$
\end{table}

This structure shows that the factor $R^{\left( l\right) }$ really has the
index structure classified by the larger group 
\begin{equation}
SO(9)_L\otimes SO(9)_R.
\end{equation}
Furthermore, the supercharge factor $2_B^{15}+2_F^{15}$ has an even larger
classification group 
\begin{equation}
SO(32)
\end{equation}
with $2_B^{15}+2_F^{15}$ corresponding to its two spinor representations%
\footnote{%
Note that $SO(32)$ contains successively $SO(9)_L\otimes SO(9)_R$,\thinspace
\thinspace $SO(10)_L\otimes SO(10)_R$,\thinspace \thinspace $SO(16)_L\otimes
SO(16)_R\,.$}. The diagonal $SO(9)$ subgroup of all these factors is the
familiar rotation group in the 10D Lorentz group $SO(9,1)$.

If there is an underlying non-perturbative 11D structure one must expect to
find SO(10) supermultiplets for both the index and base spaces at all levels 
$l.$ However, the perturbative type IIA string given above can guarantee
only SO(9) multiplets for the index space. Despite the larger classification
schemes of indices displayed above one cannot find a common SO(10) subgroup,
and hence, except for level $l=1,$ {\it the 11D structure is absent at
excited levels if only the perturbative indices are taken into account. }

One possible conclusion is that there is no 11D, and that the $l=0,1$ index
structures are just accidents. However, if one thinks of the 10D theory as a
perturbative starting point to describe the hidden 11D theory, it must be
that the 11D symmetry is broken spontaneously. This is the hint provided by
the $l=0,1$ levels discussed above. Thus, consider the scenario in which the
weak coupling type IIA string theory happens to describe a self consistent
10D part of the 11D multiplet, while the complete non-perturbative theory
has additional states, such that when combined with the perturbative string
states, they make up complete 11D multiplets. Under such an assumption the 
{\it minimal extension} of the index space of table-I that give 11D
multiplets is as follows \cite{ib11d}: At level $l_{L,R}=2$ the $9_B$ must
be extended to $10_B=9_B+1_B$ for both left/right. At level $l_{L,R}=3$ the $%
44_B$ must be extended to $54_B=44_B+9_B+1_B,$ while the $16_B$ is already a
complete $SO(10)$ representation, and so on for higher levels. The
additional states are non-perturbative states from the point of view of
string theory, but are naturally expected as part of a multiplet if there is
an underlying 11D non-perturbative theory that is spontaneously broken.

It is interesting that this {\it minimal extension} of the indices has a
definite pattern. Namely, in order to get complete SO(10) representations at
level $l_{L,R},$ we need to add {\it new states} whose indices are
isomorphic to the indices of all previous levels. Then the following set of
indices form complete SO(10)$_{L,R}$ multiplets separately for L/R movers 
\begin{equation}
\begin{array}{l}
|l_{L,R}>_p\oplus |l_{L,R}-1>_{np}\oplus |l_{L,R}-2>_{np}\oplus \\ 
\cdots \oplus |l_{L,R}=2>_{np}\oplus |l_{L,R}=1>_{np}
\end{array}
\label{11dlorentz}
\end{equation}
as can be verified by examining Table 1. The subscript $p$ describes the
perturbative states listed in Table 1, while the subscript $np$ describes
the non-perturbative states that need to be added, but whose indices are
isomorphic to those in Table 1. It is not clear what this pattern indicates,
although some suggestions appear in \cite{ib11d}. Although verified
explicitly up to $l=5,$ this pattern is conjectured to be true at all levels.

The additional new states were conjectured on the basis of 11D. However, in
recent work with S. Yankielowicz \cite{ibyank} we found out that they are 
{\it required} on the basis of U-duality in the compactified theory, as
explained below.

There may be additional, purely non-perturbative, complete 11D multiplets,
whose states are not connected by Lorentz transformations to the
perturbative string states in Table 1. Some such possibilities are mentioned
in \cite{ib11d}, but there may be others as well. By analysing the
compactified string theory and using U-duality we may find U-duality
transformations that connect different 11D multiplets to each other. Indeed
there are signs of such phenomena in our recent work \cite{ibyank}.

\section{Compactifications, U-duality and 11D}

\subsection{The states}

In the toroidaly compactified type II string on $R^d\otimes T^c,$ with $%
d+c=10,$ the perturbative vacuum state has Kaluza-Klein (KK) and winding
numbers, and is also labelled by the $2_B^7+2_F^7$ dimensional Clifford
vacuum. The closed string condition $L_0=\bar{L}_0$ can be satisfied without
requiring equal excitation levels $l_{L,R}$ for left/right movers. Hence the
perturbative states are 
\begin{equation}
\begin{array}{l}
({\rm {Bose}\oplus {Fermi\ oscillators})_L^{\left( l_L\right) }} \\ 
\times ({\rm {Bose}\oplus {Fermi\ oscillators})_R^{\left( l_R\right) }} \\ 
\times \,\,|vac,\,\,p^\mu ;\vec{m},\vec{n}>
\end{array}
\label{pertst1}
\end{equation}
where the $c$-dimensional vectors $\left( \vec{m},\vec{n}\right) $ are the
KK and winding numbers that label the ``{\it perturbative base}''. These
quantum numbers satisfy the relations 
\begin{eqnarray}
l_L+\frac 12\vec{p}_L^2 &=&l_R+\frac 12\vec{p}_R^2=M_d^2  \label{mass} \\
\vec{p}_R^2-\vec{p}_L^2 &=&\vec{m}\cdot \vec{n}=l_L-l_R  \nonumber
\end{eqnarray}
where $\vec{p}_{L,R}$ depend as usual \cite{t-review} on $\left( \vec{m},%
\vec{n}\right) $ and ($G_{ij},B_{ij})$ that parametrize the torus $T^c,$
while $M_d$ is the mass in $d$-dimensions $M_d^2=p_\mu ^2.$ By using the
same methods as \cite{ib11d} we can identify the following supermultiplet ``%
{\it perturbative index''} structure for the string states (\ref{pertst1})
at levels $\left( l_L,l_R\right) $ 
\begin{eqnarray}
\left( 0,0\right) &:&\left( 2_B^7+2_F^7\right) \otimes 1_L\otimes 1_R 
\nonumber \\
\left( 0,l_R\right) &:&\left( 2_B^{11}+2_F^{11}\right) \otimes 1_L\otimes
\sum_ir_{iR}^{(l_R)}  \nonumber \\
\left( l_L,0\right) &:&\left( 2_B^{11}+2_F^{11}\right) \otimes
\sum_ir_{iL}^{(l_L)}\otimes 1_R  \label{pertst} \\
\left( l_L,l_R\right) &:&\left( 2_B^{15}+2_F^{15}\right) \otimes
\sum_ir_{iL}^{(l_L)}\otimes \sum_ir_{iR}^{(l_R)}  \nonumber
\end{eqnarray}
The $2_B^{11}+2_F^{11}$ corresponds to the intermediate supermultiplet of
11D supersymmetry. The structures $\sum_ir_{iL,R}^{(l_{L,R})}$ are the same
ones listed in Table 1, but with the $SO(9)_{L,R}$ representations reduced
to representations of $SO(d-1)_{L,R}\otimes SO(c)_{L,R}.$ So, a general
perturbative string state is identified by ``index space'' and ``base
space'' in the form 
\begin{equation}
\phi _{indices}^{\left( l_Ll_R\right) }\left( base\right)  \label{states}
\end{equation}
where both the base and the indices are given through (\ref{pertst1},\ref
{pertst}) and Table 1.

The spectrum of the non-perturbative states is much richer in the
compactified theory. There are many central charges in the supersymmetry
algebra and those provide sources that couple to the NS-NS as well as R-R
gauge potentials. Therefore one finds a bewildering variety of
non-perturbative solutions of the low energy field equations as examples of
non-perturbative states that carry the non-perturbative charges. A complete
classification of all these charges, including p-brane charges will be given
elsewhere \cite{ibpsusy}. Here we concentrate on 0-branes. The base quantum
numbers are now extended to 
\begin{equation}
|vac,\,\,p^\mu ;\vec{m},\vec{n},z^I>
\end{equation}
where $z^I$ includes the $p_{11}$ of the previous section as well as many
other quantum numbers related to 0-branes. These correspond to the bosonic
scalar central operators in the SUSY algebra that can be simultaneously
diagonalized. This is the {\it non-perturbative base}.

There are two types of new non-perturbative states: those obtained by
applying oscillators on the non-perturbative base and those that cannot be
obtained in this way, but which are required to be present to form a basis
for U-duality transformations. The latter are generalizations (at fixed $%
l_{L,R})\,$ in the spirit of the extra states listed in (\ref{11dlorentz}),
but not necessarily identical (see below). So, a general state in the theory
is identified at each $l_{L,R}$ as in (\ref{states}). Both the base and the
indices have non-perturbative extensions as explained in \cite{ibyank}. The
full set of states turns out to form a basis for U-duality transformations
at each fixed value of $l_{L,R}.$ These states are not degenerate in mass,
hence the idea of a multiplet is analogous to the multiplets in a theory
with broken symmetry.

The BPS saturated states are those with either $l_L=0$ or $l_R=0.$ Even for
BPS saturated states there are the two types of non-perturbative states. In
particular the non-perturbative indices occur for $l_L\geq 2,\,\,l_R=0$ (or
interchange $L,R)$ similar to (\ref{11dlorentz}).

For the BPS saturated states one can derive an exact non-perturbative
formula for the mass by using the supersymmetry algebra with central
charges. For example for a state with non-trivial quantum numbers ($\vec{m},%
\vec{n},p_{11})$ and $l_R=0,$ $l_L=\vec{m}\cdot \vec{n},$ the mass is \cite
{ibpsusy} 
\begin{equation}
M_d^2=p_{11}^2+\frac 12\vec{p}_R^2.
\end{equation}
The presence of the non-perturbative $p_{11},$ as in (\ref{massp11}),
modifies the mass formula (\ref{mass}). Further generalizations involving
other non-trivial $z^I$ will be given elsewhere \cite{ibpsusy}. For non BPS
saturated states we cannot give an exact mass formula.

\subsection{Dualities}

In this paper I will give some examples of how U-duality acts on the
non-perturbative states (including oscillators) to connect them to
perturbative states, and how from these transformation properties one can
obtain the content of the quantum numbers for both the non-perturbative base
and the non-perturbative indices\footnote{%
After these lectures were delivered last summer, it was later understood
that some of the non-perturbative states discussed here and in \cite{ibyank}
are related to D-branes \cite{Polchinski}-\cite{wit-pol}, as explained in 
\cite{sen-u}. See further remarks at the end of section 2.3.}. Details of
these ideas \cite{ibyank} and their extensions to include p-branes appear
elsewhere \cite{ibpsusy}. I must emphasize that in this way we can consider
both BPS saturated as well as BPS non-saturated states. In \cite{ibyank} the
emphasis was on BPS non-saturated states whereas here I will discuss BPS
saturated ones.

The $T$-duality group is $T=O(c,c;Z)$ in all cases \cite{t-review}. The
non-compact $U$-groups, their maximal compact subgroups $K\subset U,$ and
the maximal compact subgroup $k$ of the T-group, 
\begin{equation}
k=O\left( c\right) _L\times O(c)_R
\end{equation}
are listed for various dimensions in Table 2 \cite{ht}.

\begin{table*}[tbp]
\caption{Duality groups and compacts subgroups.}
\label{tab: II}\centering
$
\begin{tabular}{|l|c|c|c|}
\hline
$d/c$ & $U$ & $K$ & $k$ \\ \hline
9/1 & $SL(2)\otimes SO(1,1)$ & $U(1)$ & $Z_2$ \\ \hline
8/2 & $SL(3)\otimes SL(2)$ & $SO(3)\otimes U(1)$ & $U(1)\otimes U(1)$ \\ 
\hline
7/3 & $SL(5)$ & $SO(5)$ & $SO(3)\otimes SO(3)$ \\ \hline
6/4 & $SO(5,5)$ & $SO(5)\otimes SO(5)$ & $SO(4)\otimes SO(4)$ \\ \hline
5/5 & $E_{6,6}$ & $USp(8)$ & $Sp(4)\otimes Sp(4)$ \\ \hline
4/6 & $E_{7,7}$ & $SU(8)$ & $SU(4)\otimes SU(4)$ \\ \hline
3/7 & $E_{8,8}$ & $SO(16)$ & $SO(7)\otimes SO(7)$ \\ \hline
\end{tabular}
$
\end{table*}

Since $T\subset U$ then $k\subset K.$ It is understood that these groups are
continuous in supergravity but only their discrete version can hold in
string theory.

The string states involved in the T-duality transformations are not all
degenerate in mass. Therefore, T-duality must be regarded as the analog of a
spontaneously broken symmetry, and the string states must come in complete
multiplets despite the broken nature of the symmetry. It is well known that $%
T=O(c,c;Z)$ acts linearly on the the $2c$ dimensional vector $\left( \vec{m},%
\vec{n}\right) $ \cite{t-review}. However it is important to realize that it
also acts on the indices in definite representations. The action of $T$ on
the indices is an induced $k$-transformation that depends not only on {\it %
all} the parameters in $T$ $\,$but also on the background $c\times c$
matrices ($G_{ij},B_{ij}$) that define the tori $T^c$ \cite{ibyank}. Since
the states in the previous section are all in $k=O(c)_L\times O(c)_R$
multiplets, the $T$-duality transformations do not mix perturbative states
with non-perturbative states.

A U-multiplet contains both perturbative as well as non-perturbative
T-multiplets. Like the $T$-duality transformations, the U-duality
transformations act {\it separately} on the base and the indices of the
states described by (\ref{states}) {\it without mixing index and base spaces}%
. The action on the base quantum numbers $(\vec{m},\vec{n},z_I)$ is in
linear representations (the repesentations are explicitly known, see e.g. 
\cite{ibyank}). The action on index space is an{\it \ induced
field-dependent gauge transformation in the maximal compact subgroup }$K$,
whose only free parameters are the global parameters in $U.$ This $\left(
U,K\right) \,$ structure extends the situation with the $\left( T,k\right) $
structure of the T-duality transformations described in the previous
paragraph. The logical/mathematical basis for this structure is more fully
explained in \cite{ibyank}. The bottom line is that in order to have
U-duality multiplets, in addition to the non-perturbative base, {\it the
``indices'' on the fields in (}\ref{states}{\it ) must be extended to form
complete }$K${\it -multiplets}.

By knowing the structure of a U-multiplet we can therefore predict
algebraically the quantum numbers of the non-perturbative states by
extending the quantum numbers of the known perturbative states given in (\ref
{pertst}). The prediction of these non-perturbative quantum numbers is one
of the immediate outcomes of our approach. In addition, our formulation
sheds some light and raises some questions on other non-perturbative aspects
of the underlying theory (such as 11 dimensional structure, see below), and
also hopefully provide hints for its non-perturbative formulation.

\subsection{An example}

It is very easy to analyze the case $\left( d,c\right) =\left( 6,\,4\right) $
so we present it here as an illustration$.$ In this case the spin group is $%
SO(5)$ and there are $4$ internal dimensions$.$ The duality groups and index
spaces follow from Tables 1,2 and (\ref{pertst}). The relevant information
is summarized by 
\begin{equation}
\begin{array}{l}
U=SO(5,5),\quad K=SO\left( 5\right) \otimes SO\left( 5\right) \\ 
T=SO(4,4),\quad k=SO\left( 4\right) _L\otimes SO\left( 4\right) _R \\ 
l_{L,R}=1:\quad \left( \sum_ir_i^{(l_{L,R})}\right) _{L,R}=1_{L,R} \\ 
l_{L,R}=2:\quad \left( \sum_ir_i^{(l_{L,R})}\right) _{L,R}=9_{L,R} \\ 
\quad \quad \quad \quad \quad =5_{L,R}^{space}\oplus 4_{L,R}^{internal} \\ 
l_{L,R}=3:\quad etc.
\end{array}
\end{equation}
where the indices $9_{L,R}$ have been reclassified according to their space
and internal components. The reclassification is done also for the short ($%
2_B^7+2_F^7),$ intermediate ($2_B^{11}+2_F^{11})$ and long ($%
2_B^{15}+2_F^{15})\,$ supermultiplet factors (see footnote \#2). It is clear
from this form that the $k=SO\left( 4\right) _L\otimes SO\left( 4\right) _R$
structure follows directly from the separate left/right internal components,
while the spin of the state is to be obtained by {\it combining} left and
right content of the space part.

Here I will discuss an example involving BPS states which is very similar to
another discussion on non-BPS states given in \cite{ibyank}. Let us consider
the BPS saturated states $\left( l_L\neq 0,\,\,l_R=0\right) .$ The base
quantum numbers in $\phi _{indices}^{\left( l_L,0\right) }\left( base\right) 
$ form the 16 dimensional spinor representation of $U=SO(5,5)$ 
\begin{equation}
base=\left( \vec{m},\vec{n},z^I\right) =16{\rm \ \ {\ of }\ \ SO(5,5)}
\label{the16}
\end{equation}
Among these the eight quantum numbers $\left( \vec{m},\vec{n}\right) $ are
perturbative, while the remaining eight $z^I$ are non-perturbative. 0-branes
that carry these quantum numbers provide the sources for the field equations
of the 8 massless NS-NS vectors and the 8 R-R vectors respectively. The
representation content of the indices in $\phi _{indices}^{\left(
l_L,0\right) }\left( base\right) $ is 
\begin{equation}
\begin{array}{l}
indices=(2_B^{11}+2_F^{11})\times \\ 
\quad \times \left[ 
\begin{array}{l}
\left( \sum_ir_i^{(l_L)}\right) _L \\ 
+non-perturbative
\end{array}
\right]
\end{array}
\end{equation}
where $(2_B^{11}+2_F^{11})$ is interpreted as the SUSY factor. The full set
of indices must form complete $K=SO\left( 5\right) _L\otimes SO\left(
5\right) _R$ multiplets for consistency with the general U-duality
transformation. It can be shown generally that the SUSY factor does satisfy
this requirement because the supercharges themselves are representations of $%
SO(5)_{spin}\times K$ \cite{ibyank}. Therefore, the remaining factor in
brackets must be required to be complete $SO(5)_{spin}\times K$ multiplets.

At level $l_L=1$ the piece $\sum_ir_i^{(1)}=1$ is a singlet, as seen in
Table 1. Hence no additional non-perturbative indices are needed at this
level. At level $l_L=2$ the piece $\sum_ir_i^{(2)}=9_L=5_L^{space}\oplus
4_L^{internal}$ is classified under $SO(5)_{spin}\times SO\left( 4\right)
_L\otimes SO\left( 4\right) _R\,$ as 
\begin{equation}
\left( 5,\left( 0,0\right) \right) +\left( 0,\left( 4,0\right) \right) .
\end{equation}
Obviously, this is not a complete $SO(5)_{spin}\times SO\left( 5\right)
_L\otimes SO\left( 5\right) _R$ multiplet. Therefore, non-perturbative
indices must be added just in such a way as to extend the $\left( 4,0\right) 
$ of $k=SO\left( 4\right) _L\otimes SO\left( 4\right) _R$ into the $\left(
5,0\right) $ of $K=SO\left( 5\right) _L\otimes SO\left( 5\right) _R.$ That
is 
\begin{equation}
\left( 4_{int}\right) _L\rightarrow \left( 5_{int}\right) _L.  \label{4to5}
\end{equation}
This extension determines the required non-perturbative indices for this
case. Note that this amounts to extending the $9_L$ into a $10_L,$ and
similarly for right-movers 
\begin{equation}
9_{L,R}\rightarrow 10_{L,R}.  \label{9to10}
\end{equation}
This is precisely what was needed in section-1 in order to obtain
consistency with an underlying 11D theory \cite{ib11d}.

At all higher levels $l_{L,R}$ the requirement for complete $K-$multiplets
coincides precisely with the requirement of an underlying 11D theory.
Therefore the full set of indices are the same as those given in eq.(\ref
{11dlorentz}). The story is the same with the non-BPS-saturated states at
arbitrary $l_{L,R}.$ This result was found in \cite{ib11d} by assuming the
presence of hidden 11-dimensional structure in the non-perturbative type-IIA
superstring theory in 10D. In ref.\cite{ib11d} a justification for (\ref
{9to10}) could not be given. However, in \cite{ibyank} and in the present
analysis $U$-duality demands (\ref{4to5}) and therefore justifies (\ref
{9to10}), and similarly for all higher levels.

Therefore for this particular compactification on $R^6\otimes T^4,$
U-duality and 11D Lorentz representations imply each other.

When similar results were reported last summer in these lectures, D-branes
had not yet entered the duality picture. However, later a consistency check
between U-duality and D-branes was reported in \cite{sen-u}. It is of
interest to compare that analysis to ours at the time of writing these
lectures. We find complete agreement at level $l_L=1.$ But at higher levels $%
l\geq 2$ our scheme requires more states than the D-brane degeneracy
computed in \cite{sen-u}. There the states corresponding to the
non-perturbative indices were not considered, seemingly because the special
U-duality transformation considered (interchanging the two 8's in the 16 of (%
\ref{the16})) has a trivial transformation on our index space (does not go
outside of the 4$_{int}^{L,R}$). We have seen that under more general
U-transformations the extra indices are needed both for U-duality multiplets
as well as for 11D interpretation. Thus, the D-brane or other interpretation
of these extra states is currently unknown.

For $\left( d,c\right) =\left( 10,0\right) ,\left( 9,1\right) ,\left(
8,2\right) ,\left( 6,4\right) $ the analysis for $l_{L,R}=2,3,4,5$ produces
exactly the same conclusion as the 11D analysis. That is, $U$-duality
demands that the $SO(9)_L\otimes SO(9)_R$ multiplets $\sum_ir_i^{(l_{L,R})}$
should be completed to $SO(10)_L\otimes SO(10)_R$ multiplets. The minimal
completion (\ref{11dlorentz}) is sufficient in this case. Hence, in these
compactifications $U$-duality is consistent with a hidden 11D structure, and
in fact they imply each other.

On the other hand for the other values $(d,c)=\left( 7,3\right) ,\left(
5,5\right) ,\left( 4,6\right) ,\left( 3,7\right) $ the story is more
complicated. At various low levels we found that the minimal index structure
required to satisfy $U$-duality is different than the {\it minimal structure}
of 11-dimensional supersymmetry multiplets (\ref{11dlorentz}). If both
U-duality and 11D are true then there must exist an even larger set of
states such that they can be regrouped either as 11D multiplets or as
U-duality multiplets. Exposing one structure may hide the other one. In fact
we have shown how this works explicitly in an example in the case $\left(
7,3\right) $ at low levels $l_{L,R}$ \cite{ibyank}. However, it is quite
difficult to see if the required set of states can be found at all levels.

\section{Summary}

We have found that at levels $l_{L,R}=0,1$ the existing index structure for
perturbative states is all that is needed to define complete $U$-multiplets
in the form $\phi _{indices}^{\left( l_L,l_R\right) }(base)$ for all values
of $\left( d,c\right) $, and that this result directly follows from the
simplest short, intermediate and long multiplet structure of 11D space-time
supersymmetry. This is easily seen since in table-I the first entry $1_{L,R}$
is just a singlet.

At levels $l_{L,R}=0,1$ all non-perturbative aspects appear in the $%
base=\left( \vec{m},\vec{n},z^I\right) $. The base quantum numbers are the
central charges of the 11D SUSY algebra and these correpond to the 0-brane
sources that couple to the massless vector particles in supergravity
(generalizations including p-brane central charges in the SUSY algebra are
found in \cite{ibpsusy}). $U$ acts as a linear transformation on the base in
a representation that is identical to the one applied to the massless vector
fields in compactified 11D supergravity. Furthermore the indices correspond
to complete representations of $K$ and they mix with a transformation
induced by $U.$ Hence, for $l_{L,R}=0,1$ both index space and base space of
U-multiplets have firm connections to 11D, for all compactifications.

To have $U$-duality at higher levels $l_{L,R}\geq 2$ additional
non-perturbative states are needed to complete the index structure. If these
additional states are absent in the theory there is no $U$-duality in the
full theory. Assuming that U-duality is true for $l\geq 2$, our approach
provides an algebraic tool for identifying the {\it minimal}
non-perturbative states at every level once the perturbative states are
listed as in Table 1.

There seems to be a non-perturbative 11D structure lurking behind the
theory. In view of the existence of a classical membrane theory with some
promise of its consistency at the quantum level, or a related M-theory,
searching for hidden 11 dimensional structure is an interesting challenge.
There is mounting evidence that 11D is present in the 10D theory, including
the work we presented here and in \cite{ib11d}\cite{ibyank}. We have seen
that U-duality is distinct from this 11D structure, although in some cases
they appeared to imply each other. We have found cases where there is a
clash between the two if one or the other is restricted to a minimal set of
non-perturbative states. We have shown, at least in one example, that the
conflicts may be resolved by adding more non-perturbative states
(non-minimal ones \cite{ibyank}). But nevertheless this example clearly
shows that 11D and U-duality are quite distinct from each other. If they are
both true, their combined effect is quite restrictive on the
non-perturbative structure of the theory. Whether the conflict can be
resolved generally is a major question raised by our work.

\end{document}